\documentclass[final,natbib]{svjour3}
\usepackage{xspace}
\usepackage{nicefrac}
\usepackage{url}
\usepackage{amsmath}
\usepackage{amsfonts}
\usepackage{braket}
\usepackage{feynmf}
\usepackage{mathptmx}
\usepackage{endnotes}

\newcommand{\ItBit}{\emph{It from Bit}\xspace}

\newcommand{\Quote}[1]{`{#1}'\xspace}
\newcommand{\QuoteDbl}[1]{``{#1}''\xspace}
\newcommand{\ie}{i.e.\xspace}
\newcommand{\eg}{e.g.\xspace}

\newcommand{\df}{\nu}
\newcommand{\Schrodinger}{Schr{\"o}dinger\xspace}
\newcommand{\Xm}{$X$-machine\xspace}

\newcommand{\PHI}[1]{\phi_{#1}}
\newcommand{\PSI}[1]{\psi_{#1}}
\newcommand{\DPhi}[1]{\Delta_{#1}}

\newcommand{\Fixed}[1]{{#1}^\dagger}
\newcommand{\Anti}[1]{\overline{#1}}

\newcommand{\MIN}{\mathrm{min}}
\newcommand{\MAX}{\mathrm{max}}
\newcommand{\tMin}{\ensuremath{t_\MIN}}       
\newcommand{\tMax}{\ensuremath{t_\MAX}}       

\newcommand{\tPath}{\ensuremath{\delta t}}    

\newcommand{\Tuple}[1]{ \boldsymbol{(} {#1} \boldsymbol{)} } 
\newcommand{\Path}[1]{[\,#1\,]}          

\newcommand{\HA}[1]{\Braket{#1}_h}               
\newcommand{\Mod}[1]{\left\lvert{#1}\right\rvert}
\newcommand{\SqMod}[1]{\Mod{#1}^2}               

\newcommand{\Sans}[1]{\ensuremath{\mathbb{#1}}\xspace}  

\newcommand{\Cset}{\Sans{C}}                     
\newcommand{\Rset}{\Sans{R}}                     

\journalname{Natural Computing}
\begin{document}

\title{The Computational Status of Physics}
\subtitle{A Computable Formulation of Quantum Theory}

\author{Mike Stannett}
\institute{Mike Stannett
  \at Department of Computer Science, Regent Court, 211 Portobello St, Sheffield S1 4DP, UK\\
  \email{M.Stannett@dcs.shef.ac.uk}%
  }

\titlerunning{A Computational Formulation of Quantum Theory}
\authorrunning{Mike Stannett}

\maketitle

\begin{abstract}
According to the Church-Turing Thesis (CTT), effective formal behaviours can be simulated by Turing machines; this has naturally led to speculation that \emph{physical} systems can also be simulated computationally. But is this wider claim true, or do behaviours exist which are strictly \emph{hypercomputational}? Several idealised computational models are known which suggest the possibility of hypercomputation, some Newtonian, some based on cosmology, some on quantum theory. While these models' physicality is debatable, they nonetheless throw into question the validity of extending CTT to include \emph{all} physical systems. 

We consider the physicality of hypercomputational behaviour from first principles, by showing that quantum theory can be reformulated in a way that explains why physical behaviours can be regarded as `computing something' in the \emph{standard computational} state-machine sense. While this does not rule out the physicality of hypercomputation, it strongly limits the forms it can take. Our model also has physical consequences; in particular, the \emph{continuity of motion} and \emph{arrow of time} become theorems within the basic model.

\keywords{%
  Hypercomputation \and
  quantum theory \and
  theory of computation \and
  philosophy of science \and
  arrow of time \and
  discrete time \and
  natural computation
  }
\subclass{%
	68Q05 \and 
  81P10 \and 
  81P68      
  }
\PACS{%
	89.20.-a \and 
	03.67.Lx      
	}

\end{abstract}

\section{Introduction}
\label{sec:introduction}

According to the Church-Turing Thesis (CTT), all effective computational behaviours can be simulated by Turing machines \citep{Kle52}. Although CTT was proposed in the context of formal mathematical systems, it is widely accepted that it can be applied more generally; in particular, given that physical devices are routinely used for computational purposes, it is now widely assumed that all (finitely-resourced, finitely-specified) physical machine behaviours can be simulated by Turing machines. However, this extended claim\endnote
    {
     Andr\'eka \textit{et al.} (\citeyear{ANN08}) argue that the physical variant of CTT was first 
     considered as far back as the 1930s.
    }
(known in the philosophy and computer science literature as \emph{Thesis M} \citep{Gan80,Cop02}, and in physics literature as the \emph{physical Church-Turing Thesis}; see \eg \citep{Deu85,Pen90} and references therein) is not by any means a logical consequence of CTT, since it is not clear that every physical machine can meaningfully be said to `compute something' in the same sense as Turing machines. Proponents of \emph{digital physics} \citep{Wol02,Llo06,Teg08} stretch CTT still further, interpreting it to mean that \emph{all} physical behaviours (whether machine-generated or not) are Turing-simulable.

The main aim of this paper is to investigate Thesis M and its extensions in more detail. Is it actually true that all physical behaviours are necessarily computable, or are there behaviours which go beyond the Turing limit? We will show that quantum theory can be reformulated in a way that partially resolves this question, by explaining why physical behaviours can indeed \emph{always} be regarded as `computing something' in the strict state-machine sense. While our approach does not rule out the possibility of hypercomputation completely, it limits the form such hypercomputation must take.

As we recall in section \ref{sec:motivation}, this question has been debated indirectly over many decades \citep{Sta06}; but it has become prominent recently with the rise of quantum computation and digital physics. As is well known, Shor's (\citeyear{Sho94}) algorithm can factorise integers faster than any Turing program, and this already suggests that quantum theory has super-Turing potential. However, we need to distinguish carefully what we mean by `hypercomputation' in this context. Where a computational model---for example, Deutsch's (\citeyear{Deu85}) Universal Quantum Computer (UQC)---computes the same class of functions as the Turing machine, albeit potentially faster, we call it a \emph{super-Turing} model. If it is capable of computing functions which \emph{no} Turing machine can compute, we call it \emph{hypercomputational}. In particular, then, while the UQC is an apparently super-Turing model, it is well known that it is not hypercomputational, whence its implementation would not resolve the question whether hypercomputation is physically feasible.

\subsection{Layout of the paper}
\label{sec:layout-of-the-paper}

We begin in section \ref{sec:motivation} by considering briefly what is already known concerning the relationship between physics and (hyper)computation. After summarising the information-theoretic approach familiar from \ItBit, we review three known hypercomputational systems: non-collision singularities in the Newtonian $n$-body problem; the Swansea \textit{Scatter Machine Experiment} (also Newtonian); and Hogarth's cosmologically inspired family of $SAD$ computers. We then focus on quantum theory, where it is unclear whether any hypercomputational model has yet been established. The question then arises whether a new approach might be able to resolve the issue. We will show that this is indeed the case, though only to a limited extent, by deriving a first-principles reformulation of Feynman's path-integral model; we review the standard formulation briefly in section \ref{sec:standard-formulation}, and present our new formulation in section \ref{sec:finitary-formulation}.

In our version of Feynman's model, there is no such thing as a continuous trajectory. Instead, whenever a particle moves from one spacetime event to another, it does so by performing a finite sequence of `hops', where each hop takes the particle directly from one location to another, with no intervening motion. Although this seems somewhat iconoclastic, we argue that `finitary' motion of this kind is the only form of motion actually supported by observational evidence.

In section \ref{sec:computational-significance} we consider the computational significance of the model, insofar as it addresses the question whether hypercomputation is physically feasible. From a mathematical point of view it makes little difference whether we allow `hops' to move a particle backwards as well as forwards in time, and we consider both models. In each case, the motion of a particle from one location to another generates a finite state machine (technically, an extended form of FSM called an \Xm \citep{Eil74}), where the machine's states are spacetime locations, and its transition labels reflect the (classical) action associated with each hop. In unidirectional time, the regular language generated by such a machine comprises just a single word, but if we allow time to be bidirectional, the availability of loops ensures that infinite regular languages can be generated. In both cases, when the motion is interpreted as an \Xm, the function computed by the motion can be interpreted as an amplitude, and if we sum the amplitudes of all machines with a given initial and final state, we obtain the standard quantum mechanical amplitude for the particle to move from the initial to the final location. 

Section \ref{sec:conclusions} concludes our argument, and includes suggestions for further research. We note in particular that certain assumptions inherent in Feynman's original model must be regarded as \emph{provable theorems} of the model presented here; this includes both the \emph{continuity of observed motion} and the \emph{arrow of subjective time}.

\section{Motivation}
\label{sec:motivation}

In this section we review various arguments both for and against the physical feasibility of hypercomputation, and its converse, digital physics; for a more complete discussion of hypercomputational models, readers are invited to consult our earlier surveys of the field \citep{Sta03,Sta06}. The question, whether hypercomputational behaviours are physically feasible, obviously depends on ones conception of physics itself. Hypercomputational systems have been identified with respect to both relativistic and Newtonian physics. Where quantum theory is concerned, however, the situation is less clear cut.

\subsection{Digital physics}
\label{sec:models:digital-physics}

Proponents of digital physics argue that the Universe \emph{as a whole} is essentially computational, in the sense that its entire history can be viewed as the output of a digital computation \citep{Sch97}. The underlying idea appears first to have been proposed by Zuse, who suggested as early as 1967 that the Universe might be computed by a deterministic cellular automaton inhabited by \Quote{digital particles} \citep{Zus67,Zus69}. 

Wheeler's subsequent (\citeyear{Whe90}) \ItBit conception reflected his conviction that information is just as physical as mass and energy, and indeed the relationship between information and gravitation has remained central to theories of quantum gravity ever since \cite{Bek72,Bek73} realised that black holes must possess intrinsic entropy. Likewise, Hawking's observation that black holes can evaporate \citep{Haw74} forces us to ask what happens to quantum correlations that previously existed between particles on either side of the event horizon? Quantum theory appears to be inconsistent with causality in such a situation \citep{SL05}.\endnote
  {
  There is as yet no empirical evidence that Hawking radiation, the mechanism by which evaporation takes place, 
  exists in Nature. However, the final stages of a primordial micro black hole's evaporation should theoretically
  result in a burst of gamma-rays; one of the goals of the GLAST satellite, launched by NASA on 11th June 2008, 
  is to search for such flashes.
  } 

The \ItBit doctrine focusses on the relationship between observation and information. Just as observations provide information, so information can affect observations, as was graphically illustrated (at first theoretically and eventually experimentally) by Wheeler's famous \Quote{delayed-choice experiment}, a modified version of the dual-slit experiment. As is well known, if one slit in a barrier is covered over, photons passing through the apparatus behave like particles, but when both slits are opened the \Quote{particles} demonstrate interference effects. Wheeler asked what would happen if the decision to cover or uncover a slit were made \emph{after} the photon had passed through the barrier, but before the outcome were detected. In practice, the photon's behaviour reflects the decision the experimenter will eventually make, even though this decision occurs after the encounter with the barrier has taken place. This suggests that the outcome of an experiment involves an interaction between the apparatus and the observer; the results you get are in some sense changed by the questions you decide to ask; or as Wheeler put it, \QuoteDbl{Every \Quote{it} -- every particle, every field of force, even the spacetime continuum itself -- derives its function, its meaning, its very existence entirely -- even if in some contexts indirectly -- from the apparatus-elicited answers to yes-or-no questions, binary choices, bits} \citep{Hor91}.

\cite{Sch97,Sch00} has investigated a model of physics in which all possible realities are the outcomes of computations. By considering algorithmic complexity, we can examine the probability that a randomly selected universe would conform to any given set of behaviours; specific physical situations can be examined and predictions made, some of which might, in principle, be subject to experimental verification. It is important to note, however, that the type of physics this model generates is \emph{not} generally consistent with conventional wisdom. For example, because digital physics assumes that universes are inherently deterministic, Schmidhuber's model rejects the notion that beta decay is truly random. Similarly, his model suggests that experiments carried out on widely-separated, but initially entangled, particles, should display non-local algorithmic regularities, a prediction which, he notes, \Quote{runs against current mainstream trends in physics}.

A related concept is Tegmark's \emph{Mathematical Universe Hypothesis}. \cite{Teg08} notes that, if a complete Theory of Everything (TOE) exists, then the Universe must necessarily be a mathematical structure. In essence, this is because a \emph{complete} TOE should make sense to any observer, human or otherwise, whence it ought to be a formal theory devoid of \Quote{human baggage}; consequently the TOE (and hence the Universe it specifies) is a purely mathematical structure. While this argument can obviously be challenged---it is entirely possible that pure mathematics is itself a form of human baggage and that the concept \Quote{mathematical structure} has no meaning to creatures whose brains have evolved differently to our own---Tegmark shows that it entails a surprisingly wide range of consequences, but interestingly, these do \emph{not} include computability. Rather, Tegmark introduces an additional \emph{Computable Universe Hypothesis}, according to which the relations describing the Universal structure can be implemented as halting computations. This is similar to Schmidhuber's model, except that it is the relationships between objects that are deemed computable, rather than their evolution through time.

\subsection{Examples of physical hypercomputation}
\label{sec:why-reformulate}

A key feature of the digital physics models described above---as well as, e.g. Zizzi's (\citeyear{Ziz04}) loop quantum gravity model---is that the models take the assumption of an information- or computation-based universe as their \emph{starting point}, and then ask what consequences follow. This is inevitable, since the authors are ultimately interested in identifying experiments which might provide evidence in support of (or which falsify) their models. Clearly, however, if experiments are to distinguish between digital physics and \Quote{conventional wisdom}, it must first be necessary that digital physics and the standard model are not equivalent. It follows, therefore, that digital physics cannot tell us about the feasibility or otherwise of hypercomputation in \Quote{standard} quantum theory.

Unfortunately, this is precisely the question we wish to answer. Rather than invent a \emph{new} model of physics that is computational by fiat, we wish to determine whether the \emph{standard} model is computational. Our approach, which we outline in some detail in sections \ref{sec:standard-formulation} and \ref{sec:finitary-formulation}, is to reformulate (a small part of) the existing model in such a way that its computational nature becomes intuitively obvious. Before doing so, however, we should explain why this task is worth undertaking---as \cite{Zus69} put it, \QuoteDbl{Is Nature digital, analog or hybrid? And is there essentially any justification for asking such a question?}

\subsubsection{Newtonian models (and a challenge to digital physics)}
\label{sec:models:newtonian}

It is not often appreciated that standard Newtonian physics supports both super-Turing and hypercomputational behaviours, but as \cite{Xia92} has shown, the Newtonian $n$-body problem exhibits \Quote{non-collision singularities}, solutions in which massive objects can be propelled to infinity in finite time. This is particularly problematic for those models of digital physics which claim the Universe is generated by essentially local interactions, like those connecting processes in a cellular automaton, because the laws of physics are typically considered to be time-reversible. Consequently, if a particle can be propelled \emph{to} infinity in finite time, it should also be possible for a particle to arrive \emph{from} infinity in finite time. Clearly, however, there is no earliest time at which such an emerging particle first arrives in the Universe (the set of times at which the emerging particle exists does not contain its greatest lower bound). Consequently, if all objects in the Universe have finite extent and finite history, the particle's \Quote{emergence at infinity} must involve some non-local form of interaction between infinitely many of these objects. On the other hand, Xia's model depends implicitly on an idealised version of Newtonian physics, in which gravitationally bound point-masses can approach arbitrarily closely (some such idealisation is unavoidable, as the system needs to supply unbounded kinetic energy to the escaping object as it accelerates away to infinity). While this means that Xia's result doesn't actually undermine the case for digital physics in `real-world' terms, it reminds us that the situation is considerably more complicated than it might at first appear.

A recent series of investigations, reported in Beggs \textit{et al.} (\citeyear{BCLT08}), concerns a collision-based computational system called the \emph{Scatter Machine Experiment} (SME), in which a projectile is fired from a cannon at an inelastic wedge in such a way that it bounces into a detector either to one side (\emph{up}) of the apparatus or the other (\emph{down}); if the projectile hits the vertex, various scenarios can be posited. The wedge is fixed in position with its vertex at some height $x$ whose binary expansion we wish to compute. The cannon can also be moved up and down, but whereas $x$ can take any real value, we only allow the cannon to be placed at heights $u$ which can be expressed in the form $u = m/2^n$ for suitable $m$ and $n$. By repeatedly firing and then re-aligning the cannon, we can attempt to compute the binary expansion of $x$, one digit at a time. The class of sets which are decidable in polynomial time, when a certain protocol is used to run the SME, is exactly $P/poly$ (the complexity class of languages recognized by a polynomial-time Turing machine with a polynomial-bounded advice function). Since $P/poly$ is known to contain recursively undecidable languages \citep{Gol08}, it follows that the scatter machine experiment---despite its evident simplicity---is behaving in a hypercomputational way.

\subsubsection{Relativistic models}
\label{sec:models:relativistic}

The $SAD_n$ hierarchy is a family of computational models which exploit the properties of certain singularities in \emph{Malament-Hogarth} spacetimes \citep{Hog92, EN02}. These are singularities with computationally useful properties; in particular, if a test particle falls into the singularity, it experiences infinite proper time during its journey; but an outside observer sees the entire descent occurring in finite time. By exploiting such a singularity, we can easily solve the Halting Problem. For suppose we want to know whether some program $P$ halts. We set it running on a computer, and then send that computer into the singularity. From our vantage point, the entire process lasts just a finite length of time, say $T$ seconds. From the computer's point of view the descent takes forever, so if $P$ is going to halt, it will have enough time to do so. We therefore program the computer's operating system so that, if $P$ halts, a rocket is launched---this is possible for this kind of singularity---so as to arrive at some previously determined place and time, somewhat more than $T$ seconds (from our point of view) after the computer is launched. We then travel to the rendezvous point. If a rocket arrives at the scheduled time, we know that $P$ must have halted. If no rocket arrives, we know that the operating system never had cause to launch it, and we conclude that $P$ ran forever.

Hogarth refers to this hypercomputational system as an $SAD_1$ computer; it uses a standard Turing machine to run the underlying program $P$, but gains hypercomputational power from the geometrical properties of the spacetime in which that Turing machine finds itself. If we now adapt the construction to use a sequence of $SAD_1$ computers in an attempt to decide some question, the resulting ($SAD_1$ + singularity) system is called an $SAD_2$ machine, and so on. Finally, by dovetailing a sequence of machines, one from from each level of the hierarchy, and sending the whole lot into an appropriate singularity, we obtain an $AD$ machine. The $SAD_n$ machines decide precisely those first order sentences which occupy the $n^\mathrm{th}$ level of the Arithmetic Hierarchy, while the $AD$ machine can decide the whole of arithmetic \citep{Hog04}.

The physicality of Malament-Hogarth spacetime is, however, debatable, since it clearly violates the Cosmic Censorship Hypothesis \citep{Pen98}; in addition, there are various technical problems associated with the transmission of the successful-completion signal from computer to observer \citep{EN93, EN96}. However, the related approach of N{\'e}meti \textit{et al.} \citep{NA06, ND06} exploits instead the properties of slow-Kerr (\ie massive slowly-rotating uncharged) black holes, whence Cosmic Censorship is no longer an issue; they have, moreover, addressed the technical problems concerning signal transmission \citep{ANN08} (see also the related paper, elsewhere this volume).

\subsubsection{Quantum theoretical models}
Quantum mechanics is, perhaps, mankind's most impressive scientific achievement to date; it enables us to predict various physical outcomes with remarkable accuracy across a wide range of both everyday and exotic situations. In addition, as \ItBit demonstrates, there are clear parallels between quantum theory and information theory; since computation is largely seen as the study of information processing, it is not surprising that the field has proven fertile ground for researchers in both digital physics and hypercomputation theory.

One possible hypercomputational model in quantum theory is Kieu's adiabatic quantum algorithm for deciding Hilbert's Tenth problem, concerning the solution of Diophantine equations. Since this problem is known to be recursively undecidable \citep{Mat93}, Kieu's algorithm---essentially a method for searching infinite sets in finite time---must be hypercomputational. Although Kieu's claims are controversial and his algorithm has been disputed by various authors, he has sought to address these criticisms in a forthcoming paper \citep{Kie08}. For the time being, therefore, the jury is out.

\section{The Standard Path-Integral Formulation}
\label{sec:standard-formulation}

As we explained in section \ref{sec:why-reformulate}, we aim to reformulate the standard version of quantum theory from first principles in such a way that its computational aspects become essentially self-evident. We begin by recapitulating the (non-relativistic) path-integral formulation originally presented in \citep[\S\S3--4]{Fey48}; see also \citep{Fey65}. Given initial and final locations $q_I = \Tuple{x_I,t_I}$ and $q_F = \Tuple{x_F,t_F}$ (where $t_F > t_I$), the goal of the standard formulation is to determine the amplitude $\phi(q_F,q_I)$ that a particle $P$ follows a trajectory  $q_I \to q_F$ lying entirely within some prescribed non-empty open space-time region $R$. As Feynman shows, this amplitude can then be used to generate a \Schrodinger wave-equation description of the system, whence this formulation is equivalent to other standard (non-relativistic) models of quantum theory. In Section \ref{sec:finitary-formulation}, we will develop a generalised finitary formulation of the same amplitude, and show that it is equivalent to the standard path-integral formulation presented below.

For the sake of illustration, we shall assume that space is 1-dimensional, so that spatial locations can be specified by a single coordinate $x$---the extension to higher dimensions is straightforward. Furthermore, we shall assume in this paper that the region $R$ is a simple rectangle of the form $R = X \times T$, where $X$ and $T = (\tMin, \tMax)$ are non-empty open intervals in $\Rset$; this does not limit our results, because open rectangles form a base for the standard topology on $\Rset^2$, and all of our formulae are derived via integration.\endnote
  {
	Integrating over a union of disjoint rectangles is the same as summing the component integrals: 
	given any integrable function $f(x,t)$ defined on a disjoint union $R = \bigcup_{\alpha}{ R_\alpha }$, we 
	have $\int_{R}{ f } = \sum_{\alpha}{ \int_{R_\alpha}{ f } }$.%
  }

Suppose, then, that a particle $P$ is located initially at $q_I = \Tuple{x_I, t_I}$, and subsequently at $q_F = \Tuple{x_F, t_F}$, and that its trajectory from $q_I$ to $q_F$ is some continuous path lying entirely within the region $R = X \times T$. Choose some positive integer $\df$, and split the duration $\tPath = t_F - t_I$ into $\df+1$ equal segments: for $n=0, \dots, \df+1$, we define $t_n = t_I + \nicefrac{n \tPath}{(\df+1)}$, so that $t_0 = t_I$ and $t_{\df+1} = t_F$. We write $x_0, \dots, x_{\df+1}$ for the corresponding spatial locations, and define $q_n = \Tuple{x_n, t_n}$. While each of the values $x_n$ can vary from path to path, the values $t_n$ are fixed. To distinguish this situation from the situation below (where $t_n$ is allowed to vary), we shall typically write $\Fixed{q} = \Tuple{x, \Fixed{t}}$ for those locations $q_n$ whose associated $t_n$-value is fixed. We will also sometimes write $\Path{\Fixed{q}}$ or $\Path{\Fixed{q}_1, \dots, \Fixed{q}_{\df}}$ for the arbitrary path $q_I = \Fixed{q}_0 \to \Fixed{q}_1 \to \dots \to \Fixed{q}_\df \to \Fixed{q}_{\df+1} = q_F$. Apart from the fixed values $x_0 \equiv x_I$ and $x_{\df+1} \equiv x_F$, each of the $x_n$ is constrained only by the requirement that $x_n \in X$, whence the path $\Path{\Fixed{q}}$ has $\df$ degrees of freedom.

In classical physics, the \emph{action} associated with a path $p$ is given by $S = \int_p{ L\ dt }$, where the function $L = L(x(t),\dot{x}(t))$, the \emph{Lagrangian}, is a function of position $x$ and velocity $\dot{x}$, only. However, to form this integral we need to specify the motion of the particle in each subinterval $(\Fixed{t}_n, \Fixed{t}_{n+1})$, so we assume that $P$ follows some path $\Fixed{q}_n \to \Fixed{q}_{n+1}$ that is classically permissible. Each segment $\Fixed{q}_n \to \Fixed{q}_{n+1}$ of the path has associated classical action $S(\Fixed{q}_{n+1},\Fixed{q}_n)$, and probability amplitude $\Braket{\Fixed{q}_{n+1}|\Fixed{q}_n}$ defined for all $q$ and (subsequent) $q'$ by $\Braket{q'|q} = \exp{\left\{ i S(q',q) / \hbar \right\}}$. The action $S$ is determined by the classical \emph{Principle of Least Action}. This says that the classical path is one which minimises this action, so that $S(q',q) = \min \int_{t}^{t'}{ L\, dt }$. The total action associated with the path is $S\Path{\Fixed{q}_1, \dots, \Fixed{q}_{\df}} = \sum_n { S(\Fixed{q}_{n+1},\Fixed{q}_n) }$ and the associated amplitude is the product
$	\Braket{q_F | \Fixed{q}_\df}
	\Braket{\Fixed{q}_\df | \Fixed{q}_{\df-1}}
		\dots
	\Braket{\Fixed{q}_2 | \Fixed{q}_1}
	\Braket{\Fixed{q}_1 | q_I }
$.
Summing over all such paths now yields the composite amplitude
\begin{equation}
		\PHI{\df}(q_F,q_I) = 
		\frac{1}{A_\df} 
		\int{ \Braket{q_F | \Fixed{q}_\df} dx_\df
		\Braket{\Fixed{q}_\df | \Fixed{q}_{\df-1}} dx_{\df-1}
			\dots
		\Braket{\Fixed{q}_2 | \Fixed{q}_1} dx_1
		\Braket{\Fixed{q}_1 | q_I } }
		\label{eq:phi-n}
\end{equation}
where $A_\df$ is a normalisation factor. All that remains is to take the limit as $\df \to \infty$, subject to the assumption that the resulting path $x = x(t)$ is continuous. This gives us the required amplitude $\phi(x_F,x_I)$ that the particle travels from $q_I$ to $q_F$ by a trajectory that lies entirely\endnote
  {
	Strictly, only the internal points of the trajectory are required to lie in $R$. 
	Either (or both) of the endpoints $q_I$ and $q_F$ can lie 
	outside $R$, provided they are on its boundary.
  }
within $R$: 
\[
	\phi(q_F,q_I) 
	=
	\lim_{ \df \to \infty }
	\frac{1}{A_\df} 
	\int{ \Braket{q_F | \Fixed{q}_\df} dx_\df
	\Braket{\Fixed{q}_\df | \Fixed{q}_{\df-1}} dx_{\df-1}
		\dots
	\Braket{\Fixed{q}_2 | \Fixed{q}_1} dx_1
	\Braket{\Fixed{q}_1 | q_I } }
 \enspace .
\]
\section{A Finitary Formulation}
\label{sec:finitary-formulation}

In section \ref{sec:standard-formulation} we showed how the amplitude $\phi(q_F,q_I)$, that the particle $P$ travels from $q_I$ to $q_F$ along some path lying entirely within the non-empty open spacetime region $R = X \times T$, is given by $\phi = \lim_{\df \to \infty} \PHI{\df}$. If we now write 
\begin{equation}
	\DPhi{n} = \PHI{n} - \PHI{n-1} \enspace ,
	\label{eq:dphi}
\end{equation}
it follows from the identity $\PHI{\df} = (\PHI{\df} - \PHI{\df-1}) + \dots + (\PHI{1} - \PHI{0}) + \PHI{0}$ that
\[
	\lim_{\df \to \infty} \PHI{\df}
	=
	\lim_{\df \to \infty} \left( \PHI{0} + \sum_{n=1}^{\df}{ \DPhi{n} } \right)
	=
	\PHI{0} + \sum_{n=1}^{\infty}{ \DPhi{n} } \enspace .
\]

This replacing of a limit with a sum is a key feature of our model, since it allows us to describe a system in terms of a set of mutually distinct finite sets of observations. We can think of this sum in terms of \emph{correction factors}. For, suppose you were asked to estimate the amplitude $\phi(q_F, q_I)$ that some object or particle $P$ will be observed at $q_F$, given that it had already been observed at $q_I$ and was constrained to move within the region $R$. With no other information to hand, your best bet would be to assume that $P$ follows some action-minimising classical path, and so the estimate you give is the associated amplitude $\Braket{q_F|q_I}$. Some time later, you realise that one or more observations may have been made on the particle while it was moving from $q_I$ to $q_F$, and that this would have perturbed the amplitude. To take account of these possibilities, you add a series of correction factors to your original estimate; first you add $\DPhi{1}$ in case 1 observation had taken place, instead of the 0 observations you had originally assumed. Then you add $\DPhi{2}$ in case there were actually 2 observations, and so on. Each $\DPhi{n}$ takes into account the extra information acquired by performing $n$ observations instead of $n-1$, and since the overall estimate needs to take all of the corrections into account, we have $\phi = \PHI{0} + \sum{\DPhi{n}}$.

The simple truth, however, is that \emph{continuous motion cannot be observed}, because making an observation takes time. The best we can ever do is to make a series of distinct measurements showing us where an object was at finitely many closely-spaced instants $t_1, t_2, \dots, t_\df$ during the relocation from $q_I$ to $q_F$. The classical spirit within us then tells us to extrapolate these discrete points into a continuous curve (namely, that path which \Quote{best} joins the points). It is as if we draw the individual locations on celluloid, and then play a mental film projector to give ourselves the comfortable impression of continuous movement. But this mental film projector---represented in the standard formulation by the construction of $\lim \phi_\df$---is no part of physical observation; it represents instead an \emph{assumption} about the way the world \Quote{ought to be}. All we can truthfully say is that the object was at such and such a location $x_n$ when we observed it at time $t_n$, and was subsequently at location $x_{n+1}$ at time $t_{n+1}$. Regardless of underlying reality (about which we can say virtually nothing), the \emph{observed} universe is inherently discrete. We can ask ourselves how the motion appears if no observations are made; the composite answer, taking into account all potential observers, is given by some amplitude $\psi_0$. If we ask how it appears if precisely $\df$ observations are made during the relocation from $q_I$ to $q_F$, we get another amplitude $\psi_\df$. Since these possibilities are all mutually exclusive, and account for every possible finitely observed relocation from $q_I$ to $q_F$, the overall amplitude that the relocation happens is the sum of these amplitudes, namely some function $\psi = \sum{\PSI{\df}}$.

Although they both involve infinite sums, these two descriptions are very different, because $\PSI{n}$ tells us the amplitude for a path with a specific number of hops, while $\DPhi{n}$ describes what happens when we \emph{change} the number of hops. Nonetheless, prompted by the formal structural similarity of the equations $\phi = \PHI{0} + \sum_1^\infty \DPhi{n}$ and $\psi = \sum_0^\infty \PSI{n}$, we shall equate the two sets of terms, and attempt to find solutions. By requiring $\PSI{0} = \PHI{0}$ and $\PSI{n} = \DPhi{n}$ for positive $n$, this will ensure that the description we generate---no matter how unnatural it might appear at first sight---satisfies $\phi = \psi$, whence it describes exactly the same version of physics as the standard formulation.

The surprising feature in what follows is that the description we generate is \emph{not} unnatural. Quite the opposite. To see why, we need to remember that amplitudes are normally given in the form $\PHI{n} = \exp{\left\{i(S_1 + \dots + S_n))/\hbar\right\}}$. In very rough terms, we can think of the various $S$ values as being essentially equal, so that $\PHI{n} \approx \exp{\left\{inS/\hbar\right\}}$. When we compute $\DPhi{n}$, we are asking how $\PHI{n}$ changes when $n$ changes; in other words, we can think of $\DPhi{n}$ in fairly loose terms as a measure of $\nicefrac{d\PHI{n}}{dn}$. Again arguing loosely, we can calculate $\nicefrac{d\PHI{n}}{dn} \approx \nicefrac{iS\PHI{n}}{\hbar}$, and now it becomes clear why equating the two sets of terms works, for in essence, $\DPhi{n}$ is approximately proportional to $\PHI{n}$. Since $\PSI{n}$ is structurally similar to $\PHI{n}$, in the sense that both measure the amplitude associated with a sequence of jumps, it is not surprising to find a similar relationship holding between $\DPhi{n}$ and $\PSI{n}$. Since the equations we form will eventually include integrals with normalisation factors, these factors will effectively absorb any remaining constants of proportionality.

\subsection{Paths, Actions and Amplitudes}
\label{sec:paths}

The standard formulation assumes that each trajectory $x(t)$ is a consistently future-pointing\endnote
    {
    As explained in his 1965 Nobel Prize address, Feynman \citeyear{Fey65} subsequently described
    anti-particles as particles moving `backwards in time'. In effect, our own approach adopts this
    temporal bi-directionality, and places it centre-stage.
    }
spacetime path; this is implicit in the continuity of the representation $x \equiv x(t)$, which assigns one location to each $t$ in the interval $[t_I, t_F]$. Since our formulation rejects this assumption, we need to provide a different definition for \emph{paths}.

We shall assume the abstract existence of a clock, represented by the integer variable $\tau$, used to indicate the order in which observations occur. Each time the clock ticks, \ie for each $\tau = 0, 1, 2, \dots$, the particle is observed to exist at some space-time location $q_\tau = \Tuple{x_\tau, t_\tau}$. We call each transition $q_\tau \to q_{\tau+1}$ a \emph{hop}. A finite sequence of consecutive hops $q_0 \to \dots \to q_{\df+1}$ constitutes a \emph{path}. As before, we take $q_0 = \Tuple{x_I,t_I}$ and $q_{\df+1} = \Tuple{x_F,t_F}$, and consider the properties of an arbitrary path from $q_I$ to $q_F$ via $\df$ intermediate points, all of which are required to lie in the prescribed space-time region $R = X \times T$.

We again write $\Path{q_1, \dots, q_\df}$ for the path $q_I \to q_1 \to \dots \to q_\df \to q_F$.  However, whereas the intervals $t_{n+1} - t_n$ were formerly fixed to have identical duration $\nicefrac{\tPath}{(\df+1)}$, there is no constraint on the temporal separation $t_{\tau+1} - t_{\tau}$ in the finitary formulation; the path $q_0 \to \dots \to q_{\df+1}$ therefore has $2\df$ degrees of freedom, or \emph{twice} the number in the standard formulation. Notice that we now write $q_n$ rather than $\Fixed{q}_n$, to show that the value $t_n$ is no longer fixed.

What is not clear at this stage is whether hops need necessarily always be future-pointing. The standard formulation forces this on us through its assumption that some continuous motion $t \mapsto x(t)$ is being observed, but this assumption is no longer relevant. We shall therefore describe two finitary formulations, one in which hops are unidirectional in time, and one in which space and time are treated symmetrically, in that hops can move both forwards and backwards in time as well as space. Both models are related to computation theory, but the second is by far the more interesting, both from a computational, and a physical, point of view. The mathematical distinction between the two models is minor. If time is unidirectional into the future, then $t_{\tau+1}$ must lie in the range $t_\tau < t_{\tau+1} \leq \tMax$. Otherwise, it can take any value in $T$.

In the standard formulation, any unobserved motion from one observation to the next is assumed to be classical, and its amplitude is determined by minimising the classical action $S$. Since we no longer assume that any such motion exists, we shall simply assume that each hop $q \to q'$ has a \emph{hop amplitude}, denoted $\HA{q'|q}$, and that this amplitude (when it is non-zero) is associated with an abstract \emph{hop action}, denoted $s_h(q', q)$, by the formula $\HA{q'|q}  = e^{i s_h(q', q) / \hbar}$. One of our tasks will be to identify the function $s_h$.

The amplitude associated with the path $\Path{q_1, \dots q_\df}$ is defined, as usual, to be the product $\HA{q_F|q_\df} \times \dots \times \HA{q_1|q_I}$. The amplitude computed by summing over all paths of this length will be denoted $\PSI{n}$, so that the overall \emph{finitary amplitude} that the particle moves from $q_I$ to $q_F$ along a sequence of hops lying entirely within $R$ is just $\psi(q_F, q_I) = \sum_{n=0}^{\infty}{ \PSI{n} }$.

\subsection{The Finitary Equations}

Consider again the formulae giving the amplitude that a particle $P$ follows a path from $q_I$ to $q_F$ that lies entirely within the region $R$, \emph{subject to the assumption} that $q_F$ occurs later than $q_I$---the standard formulation isn't defined when this isn't the case. We can write these in the form
\begin{align}
	\phi &= \PHI{0} + \sum_{n=1}^{\infty}{ \DPhi{n} } \label{eq:phi-rec} \\
	\psi &= \PSI{0} + \sum_{n=1}^{\infty}{ \PSI{n}  } \label{eq:psi-rec}
\end{align}
whence it is clear that one particular solution can be obtained by solving the infinite family of equations
\begin{align}
	\PSI{0} &= \PHI{0} \label{eq:base} \\
	\PSI{n} &= \PHI{n} - \PHI{n-1} \quad \text{ (\ie $\PSI{n} = \DPhi{n}$) \quad  for $n > 0$ } \label{eq:step}
\end{align}
to find the hop-action $s_h$. Since the terms $\PHI{n}$ and $A_n$ are those of the standard formulation, we shall henceforth assume that $S$, $\PHI{n}$, $\DPhi{n}$ and $A_n$ are all \emph{known functions} in what follows.

\subsection{Solving the Equations}

As usual, we shall assume that $q_F$ occurs later than $q_I$ (so that $\PHI{n} = \PHI{n}(q_F,q_I)$ is defined for each $n$). We shall be careful to distinguish locations $\Fixed{q} = \Tuple{x, \Fixed{t}}$ for which the time of observation is fixed in the standard formulation, from those of the form $q = \Tuple{x,t}$ used in the finitary version, for which the value of $t$ is variable. Note first that (\ref{eq:phi-n}) can be rewritten to give us a recursive definition of $\PHI{\df}$, viz.
\begin{equation}
	\begin{aligned}
		\PHI{\df}&(q_F,q_I)
		= 
			\frac{1}{A_\df} 
			\int{
				\Braket{q_F | \Fixed{q}_\df} dx_\df
				\Braket{\Fixed{q}_\df | \Fixed{q}_{\df-1}} dx_{\df-1}
					\dots
				\Braket{\Fixed{q}_2 | \Fixed{q}_1} dx_1
				\Braket{\Fixed{q}_1 | q_I } 
			} \\
		&= 
			\frac{A_{\df-1}}{A_\df} 
			\int{ 
				\Braket{q_F | \Fixed{q}_\df} dx_\df
				\frac{1}{A_{\df-1}}
				\int{
					\Braket{\Fixed{q}_\df | \Fixed{q}_{\df-1}} dx_{\df-1}
						\dots
					\Braket{\Fixed{q}_2 | \Fixed{q}_1} dx_1
					\Braket{\Fixed{q}_1 | q_I } 
				}
			} \\
		&= 
			\frac{A_{\df-1}}{A_\df} 
			\int{ 
				\Braket{q_F | \Fixed{q}_\df}
				\PHI{\df-1}(\Fixed{q}_\df, q_I) \; dx_\df
			} \\
	\end{aligned}
	\label{eq:phi-int}
\end{equation}
and an identical derivation gives $\PSI{\df}$ in the form
\begin{equation}
		\PSI{\df}(q_F,q_I)
		=
		\frac{B_{\df-1}}{B_\df} \int_X{ \int_{T'}{ \HA{q_F|q_\df} \PSI{\df-1}(q_\df,q_I) \; dt_\df \; dx_\df }}
		\label{eq:psi-n}
\end{equation}
where the $B_n$ are normalisation factors, and the integration range $T'$ depends on whether we allow hops to jump backwards in time, or insist instead that they move only forwards (we consider the two cases separately, below).

Using (\ref{eq:phi-int}) to substitute for $\PHI{\df}$ in the definition (\ref{eq:dphi}) of $\DPhi{n}$ gives
\[
	\begin{aligned}
		\DPhi{\df}(q_F,q_I) 
		&= \PHI{\df}(q_F,q_I) - \PHI{\df-1}(q_F,q_I) \\
		&=  \left[ 
					\frac{A_{\df-1}}{A_\df} 
					\int{ 
						\Braket{q_F | \Fixed{q}_\df}
						\PHI{\df-1}(\Fixed{q}_\df, q_I) \; dx_\df
					}
				\right]
				 - \PHI{\df-1}(q_F,q_I)
				 \enspace .
	\end{aligned}
\]

The case $\df = 0$ is worth noting in detail. The amplitudes $\PHI{0}(q_F,q_I)$ and $\PSI{0}(q_F,q_I)$ describe the situation in which $P$ moves from $q_F$ to $q_I$ without being observed. In the standard formulation, it is assumed in such circumstances that $P$ follows some classical path for which the action $S$ is minimal, while in the finitary formulation we assume that the particle \emph{hops} directly from $q_I$ to $q_F$. The amplitudes for these behaviours are $\Braket{q_F|q_I}$ and $\HA{q_F|q_I}$, respectively. However, we need to remember that $\PHI{0}$ and $\PSI{0}$ are defined in terms of their contribution to the \emph{overall} amplitudes $\phi$ and $\psi$; it is important, therefore, to include the relevant normalisation factors. We therefore define, in accordance with (\ref{eq:phi-n}), (\ref{eq:phi-rec}), (\ref{eq:psi-rec}) and (\ref{eq:psi-n}),
\[
		\PHI{0}(q_F,q_I) = \frac{1}{A_0} \Braket{q_F|q_I} 
		\qquad \text{ and } \qquad
		\PSI{0}(q_F,q_I) = \frac{1}{B_0} \HA{q_F|q_I} \enspace ,
\]
so that, whenever $q_F$ occurs later than $q_I$,
\begin{equation}
	\HA{q_F|q_I}  = \sigma \Braket{q_F|q_I}
	\label{eq:base-HA}
\end{equation}
 where
\[
		\sigma = B_0 / A_0 \enspace .
\]
Taking principal logarithms on both sides of (\ref{eq:base-HA}) now gives
\[
	s_h(q_F, q_I) = S(q_F, q_I) -  i \hbar \log \sigma
\]
and if we assume that $s_h$ should be real-valued (the classical action $S$ is always real-valued), then $\log \sigma$ must be a real multiple of $i$, say $\sigma = e^{i \rho}$ where $\rho \in \Rset$, whence $\SqMod{\sigma} = 1$. Consequently, $\SqMod{\HA{q_F|q_I}} = \SqMod{\HA{q_F|q_I}}$, and the two formulations assign the same standard and finitary probabilities to the relocation $q_I \to q_F$, whenever this is unobserved and future-directed. Moreover, since
\[
	s_h(q_F, q_I) = S(q_F, q_I) +  \rho \hbar
\]
we see that our earlier intuition is essentially confirmed: the hop-action $s_h$ (the best estimate of the path-amplitude, given that no observations will be made) is just the classical action $S$, though possibly re-scaled by the addition of a constant action of size $\rho\hbar$ (which we can think of as a kind of `zero-point' action). For the purposes of this paper, the values of $\rho$ and $\sigma = e^{i\rho}$ are essentially arbitrary; we shall leave $\rho$ (and hence $\sigma$) an undetermined parameter of the model, in terms of which 
\begin{equation}
	B_0           = \sigma A_0  \label{eq:B-0}
\end{equation}
and
\begin{equation}
	s_h(q_F, q_I) = S(q_F, q_I) + \rho \hbar \quad \text{ if $q_F$ occurs after $q_I$ . } \label{eq:sh-forward}
\end{equation}
The physical significance of $\rho$ is discussed briefly in Section \ref{sec:bidirectional-model}, in relation to \emph{null-hops}.

\subsection{The Unidirectional Model}
\label{ref:unidirectional-model}

If we wish to allow only future-pointing hops---we shall call this the \emph{unidirectional} model---there is little left to do. We know from (\ref{eq:base}) and (\ref{eq:step}) that each function $\PSI{n}$ is defined in terms of the known functions $\PHI{0}$ and $\DPhi{n}$. It only remains to identify the hop amplitude $s_h$ and the normalisation factors $B_n$. As explained above, our solutions will be given in terms of the undetermined phase parameter $\sigma$.

Since the side-condition on (\ref{eq:sh-forward}) is satisfied, the hop amplitude is given in terms of the classical action by the formula $\HA{q'|q} = \sigma \Braket{q'|q} = \sigma \exp\{ i S(q',q) / \hbar \}$, whenever $q'$ follows $q$. 

To find the normalisation factors, we note first that (\ref{eq:B-0}) gives us the value $B_0 = \sigma A_0$ directly. Next, when $\df > 0$, we observe that, since $t_{\df}$ must come after $t_{\df-1}$, the range $T'$ in (\ref{eq:psi-n}) is the interval $(t_{\df-1},t_F)$. Consequently,
\begin{equation}
	\begin{aligned}
			\PSI{\df}(q_F,q_I)
			&= \frac{B_{\df-1}}{B_\df} \int_X{ \int_{t_{\df-1}}^{t_F}{ \HA{q_F|q_\df} \PSI{\df-1}(q_\df,q_I) \; dt_\df \; dx_\df }} \\
			&= \frac{\sigma B_{\df-1}}{B_\df} \int_X{ \int_{t_{\df-1}}^{t_F}{ \Braket{q_F|q_\df} \PSI{\df-1}(q_\df,q_I) \; dt_\df \; dx_\df }}
			\enspace .
	\end{aligned}
	\label{eq:psi-step-soln}
\end{equation}

When $\df = 1$, (\ref{eq:psi-step-soln}) can be rewritten
\[
\begin{aligned}
	\PSI{1}(q_F,q_I)
	&= \frac{\sigma B_0}{B_1} \int_X{ \int_{t_I}^{t_F}{ \Braket{q_F|q_1} \PSI{0}(q_1,q_I) \; dt_1 \; dx_1 }} \\
	&= \frac{\sigma B_0}{B_1} \int_X{ \int_{t_I}^{t_F}{ \Braket{q_F|q_1} \frac{1}{B_0}\HA{q_1|q_I} \; dt_1 \; dx_1 }} \\
	&= \frac{\sigma^2 }{B_1} \int_X{ \int_{t_I}^{t_F}{ \Braket{q_F|q_1} \Braket{q_1|q_I} \; dt_1 \; dx_1 }}
\end{aligned}
\]
and, since $\PSI{1} = \DPhi{1}$, this gives us
\[
	B_1 = \left(\frac{ \int_X{ \int_{t_I}^{t_F}{ \Braket{q_F|q_1} \Braket{q_1|q_I} \; dt_1 \; dx_1 }} }{ \DPhi{1}(q_F,q_I) }\right) \sigma^2 \enspace .
\]

Finally, for $\df > 1$, (\ref{eq:psi-step-soln}) becomes
\[
\begin{aligned}
  \DPhi{\df}(q_F,q_I) 
  &=\PSI{\df}(q_F,q_I) \\
  &= \frac{\sigma B_{\df-1}}{B_\df} \int_X{ \int_{t_{\df-1}}^{t_F}{ \Braket{q_F|q_\df} \PSI{\df-1}(q_\df,q_I) \; dt_\df \; dx_\df }} \\
  &= \frac{\sigma B_{\df-1}}{B_\df} \int_X{ \int_{t_{\df-1}}^{t_F}{ \Braket{q_F|q_\df} \DPhi{\df-1}(q_\df,q_I) \; dt_\df \; dx_\df }} \\
\end{aligned}
\]
and hence $B_\df$ can be defined recursively, as
\[
 B_\df = \frac{\sigma B_{\df-1}}{\DPhi{\df}(q_F,q_I)} \int_X{ \int_{t_{\df-1}}^{t_F}{ \Braket{q_F|q_\df} \DPhi{\df-1}(q_\df,q_I) \; dt_\df \; dx_\df }}
 \enspace .
\]

\subsection{The Bidirectional Model}
\label{sec:bidirectional-model}

Far more interesting is the case where hops are allowed to jump backwards as well as forwards in time. It is important to note that the derivation of $B_{\df}$ given above for the unidirectional model no longer works, because it relies on using (\ref{eq:base-HA}) to replace $\HA{q_F|q_\df}$ with $\sigma \Braket{q_F|q_\df}$, and on (\ref{eq:step}) to replace $\PSI{n+1}(q_\df, q_I)$ with $\DPhi{n+1}(q_\df,q_I)$. But our use of (\ref{eq:base-HA}) assumes that $q_F$ occurs after $q_\df$, and that of (\ref{eq:step}) that $q_\df$ comes after $q_I$, and neither assumption is generally valid in the bidirectional model. Consequently, before we can make progress, we need to decide how $\HA{q'|q}$ should be defined when the hop $q \to q'$ moves \emph{backwards} in time.

\begin{figure}[!htb]
\centering
\parbox{.8\linewidth}{
\centering
\begin{fmffile}{epscatter}
\begin{fmfgraph*}(100,50)
	\fmfleft{i1,i2} \fmfright{o1,o2}
	\fmf{photon}{i2,v1}
	\fmf{photon}{v2,o2}
	\fmfdot{v1,v2}
	\fmf{electron}{i1,v1,v2,o1}
\end{fmfgraph*}
\end{fmffile}
\caption{Anti-matter can be thought of as matter moving backwards in time.
  A particle arrives at bottom left, and the corresponding antiparticle (shown
  as usual with the arrow reversed) at bottom right; they annihilate to produce
  two gamma rays, emitted top left and top right. Time advances up the page.}
\label{fig:feyn}
}
\end{figure}
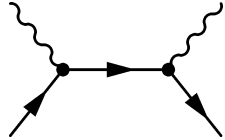

To address this problem, we recall the standard interpretation of \emph{anti-matter} as \Quote{matter moving backwards in time}. For example, the Feynman diagram in Figure \ref{fig:feyn} shows how the annihilation of \eg an electron and a positron (its antiparticle) to form two photons can be interpreted instead as showing an electron that moves forward in time, interacts with the photons, and then returns into the past.

Accordingly, whenever we are presented with a backwards hop by the particle $P$, we re-interpret it as a \emph{forwards} hop by the appropriate anti-particle, $\Anti{P}$. Writing $\Anti{S}$ for the classical action associated with the antiparticle $\Anti{P}$, we therefore define
\begin{align}
	s_h(q_F,q_I) &=
	\begin{cases}
		\rho \hbar + S(q_F, q_I) & \text{ if $q_I$ is earlier than $q_F$, and } \\
		\rho \hbar + \Anti{S}(q_I, q_F) & \text{ if $q_I$ is later than $q_F$. }
	\end{cases}
	\label{eq:backwards-hops}
\end{align}
It is tempting to assume that $\Anti{S}$ is just the negative of $S$, but this need not be the case. For example, since photons are their own anti-particles, they would require $\Anti{S} = S$. Or consider an electron moving in both an electric and a gravitational field. If we replaced it with a positron, the electric forces would reverse, but the gravitational forces would remain unchanged, and the overall change in action would reflect both effects.

\paragraph{Spatial hops - the physical meaning of $\sigma$.}
What about purely spatial hops that move the particle $P$ sideways in space, without changing its temporal coordinate? There are two cases to consider. If $q_F = q_I$, the particle has not actually moved, and the classical solution $S(q,q) = 0$ holds valid. Consequently, we can simply extend our existing solution by defining $s_h(q,q) = \rho \hbar$, or $\HA{q|q} = \sigma$. This, then, explains the physical significance of $\sigma$---it is the amplitude associated with the \emph{null hop}, \ie that hop which leaves the particle in its original location from one observation to the next.

If $q_F$ and $q_I$ differ in their $x$ (but not their $t$) values, we shall simply take $\HA{q_F|q_I} = 0$; \ie we ban all such hops (this definition is, of course, purely arbitrary, and other definitions may be more appropriate in regard to other investigations\endnote
  {
	For example, suppose we know (from wave-equation methods, say) that $P$ has amplitude $\eta(x)$ to be 
	at location $\Fixed{x} = \Tuple{x,\Fixed{t}}$, for each $x \in X$. A more intuitive solution might then be to take
	$\HA{\Fixed{x}|\Fixed{y}} = {\eta(\Fixed{x})}/{\eta(\Fixed{y})}$. This gives $\HA{\Fixed{x}|\Fixed{x}} = 1$ in
	agreement with the \Quote{classical amplitude}, but also provides information about the relative amplitudes of all 
	other spatial locations at time $\Fixed{t}$.
  };
but for our current purposes the specific choice of purely spatial hop action makes little difference, because the paths in question contribute nothing to the integrals we shall be constructing). This doesn't mean, of course, that a path cannot be found from $q_I$ to a simultaneous location $q_F$---it can, via any past or future location---but that more than one hop is required to complete the journey. Indeed, the possibility of purely spatial relocations is highly significant, since one could interpret them as explaining quantum uncertainty: one cannot say definitely where a particle is at any given time $t$, precisely \emph{because} it is able to relocate from one location to another, with no overall change in $t$.

\paragraph{Solving the Equations.}
As before, we know from (\ref{eq:base}) and (\ref{eq:step}) that each function $\PSI{n}$ is defined in terms of the known functions $\PHI{0}$ and $\DPhi{n}$, and it remains to identify the hop amplitude $s_h$ and the normalisation factors $B_n$. Once again, our solutions will be given in terms of the undetermined phase parameter $\sigma$. As always, we assume that $t_I < t_F$, although we allow individual hops to move backwards through time.

To define the hop amplitude, we appeal to (\ref{eq:backwards-hops}), and the relationship $\HA{q'|q} =$ $e^{ i s_h(q', q) / \hbar }$. Taken together with our discussion of spatial hops, these allow us to define $s_h$ fully:
\[
	\HA{q_F|q_I} =
	\begin{cases}
		\sigma \; \Anti{\Braket{q_I|q_F}} & \text{ if $q_F$ is earlier than $q_I$, } \\
		\sigma \; \Braket{q_F|q_I}        & \text{ if $q_F$ is later than $q_I$ } \\
		\sigma                            & \text{ if $q_F = q_I$, and } \\
		0                                 & \text{ otherwise. }
	\end{cases}
\]
where $\Anti{\Braket{q_I|q_F}} = \exp{\{ i \Anti{S}(q_I, q_F) / \hbar \}}$ is the \Quote{classical amplitude} associated with the antiparticle. This idea extends throughout the functions defined in this section; for example, when $q'$ is earlier than $q$, we write $\Anti{\PSI{n}}(q,q')$ for the amplitude that the antiparticle follows some path $q' \to q$ lying entirely within $R$. We will see below that the amplitude functions $\PSI{n}(q',q)$ and $\Anti{\PSI{n}}(q',q)$ are, as one would expect, related to one another in a mutually recursive way.

Now we consider the normalisation constants $B_n$. We already know that $B_0 = \sigma A_0$, so we consider the case when $n > 0$. Because hops are allowed to move in both directions through time, the integration range $T'$ in (\ref{eq:psi-n}) is the whole of $T$. Consequently, (\ref{eq:psi-n}) becomes
\[
	\PSI{\df}(q_F,q_I)
			= \frac{B_{\df-1}}{B_\df} \int_X{ \int_T{ \HA{q_F|q_\df} \PSI{\df-1}(q_\df,q_I) \; dt_\df \; dx_\df }} \enspace .
\]
The integral over $T$ splits into three parts, depending on the value of $t_\df$ relative to $t_I$ and $t_F$. We have
\begin{equation}
	\begin{aligned}
		\PSI{\df}(q_F,q_I)
				 &= \frac{B_{\df-1}}{B_\df} 
				      \int_X{ \int_T{ \HA{q_F|q_\df} \PSI{\df-1}(q_\df,q_I) \; dt_\df }\; dx_\df }  \\
				 &= \frac{B_{\df-1}}{B_\df}
					 		\int_X{ \left[ I_L(x_\df) + I_M(x_\df) + I_R(x_\df) \right] dx_\df }
	\end{aligned}
	\label{eq:psi-bi}
\end{equation}
where $I_L(x_\df)$ is the integral over $[\tMin, t_I]$, $I_M(x_\df)$ over $[t_I, t_F]$ and $I_R(x_\df)$ over $[t_F, \tMax]$.

When $\df=1$, (\ref{eq:psi-bi}) becomes
\[
	\PSI{1}(q_F,q_I) = \frac{B_0}{B_1} \int_X{ \left[ I_L(x_1) + I_M(x_1) + I_R(x_1) \right] dx_1 }
\]
and the integrals $I_L$, $I_M$ and $I_R$ are defined by
\[
	\begin{aligned}
		I_L(x_1) &= \sigma \int_{\tMin}^{t_I}{ \HA{q_F|q_1} \PSI{0}(q_1,q_I) \; dt_1  }
		         =&& \frac{\sigma^2}{B_0} \int_{\tMin}^{t_I}{ \Braket{q_F|q_1} \Anti{\Braket{q_I|q_1}} \; dt_1  } \\
		I_M(x_1) &= \sigma \int_{t_I}^{t_F}  { \HA{q_F|q_1} \PSI{0}(q_1,q_I) \; dt_1  }
		         =&& \frac{\sigma^2}{B_0} \int_{t_I}^{t_F}  { \Braket{q_F|q_1} \Braket{q_1|q_I} \; dt_1  } \\
		I_R(x_1) &= \sigma \int_{t_F}^{\tMax}{ \HA{q_F|q_1} \PSI{0}(q_1,q_I) \; dt_1  }
		         =&& \frac{\sigma^2}{B_0} \int_{t_F}^{\tMax}{ \Anti{\Braket{q_1|q_F}} \Braket{q_1|q_I} \; dt_1  } \enspace .
	\end{aligned}
\]
Thus $I_L(x_1) + I_M(x_1) + I_R(x_1) =$
\[
  \frac{\sigma^2}{B_0} \left[ 
					\int_{\tMin}^{t_I}{ \Braket{q_F|q_1}        \Anti{\Braket{q_I|q_1}}  } 
				+ \int_{t_I}^{t_F}  { \Braket{q_F|q_1}        \Braket{q_1|q_I}         }
				+ \int_{t_F}^{\tMax}{ \Anti{\Braket{q_1|q_F}} \Braket{q_1|q_I}         }
				\right]
\]
and $\PSI{1}(q_F,q_I)$ equals
\[
	\frac{\sigma^2}{B_1} \left[ 
				\int_{\tMin}^{t_I}{ \Braket{q_F|q_1}  \Anti{\Braket{q_I|q_1} }  }
			+ \int_{t_I}^{t_F}  { \Braket{q_F|q_1}        \Braket{q_1|q_I} }
			+ \int_{t_F}^{\tMax}{ \Anti{\Braket{q_1|q_F}} \Braket{q_1|q_I} }
			\right] \enspace .
\]
On the other hand, (\ref{eq:dphi}) tells us that $\PSI{1} = \DPhi{1}$, and so $B_1$ equals
\[
	\begin{aligned}
		\frac{\sigma^2}{ \DPhi{1}(q_F,q_I) } &\times  \\
		\left[ 
				\int_{\tMin}^{t_I} \right. & \left. { \Braket{q_F|q_1} \Anti{\Braket{q_I|q_1}} } 
			+ \int_{t_I}^{t_F}  { \Braket{q_F|q_1}        \Braket{q_1|q_I}  }
			+ \int_{t_F}^{\tMax}{ \Anti{\Braket{q_1|q_F}} \Braket{q_1|q_I}  }
			\right]
		  \enspace .
	\end{aligned}
\]
Finally, when $\df > 1$, the integrals $I_L$, $I_M$ and $I_R$ are given by
\begin{itemize}
\item 
  $ I_L(x_\df) = \sigma \int_{\tMin}^{t_I}{ \Braket{q_F|q_\df}        \Anti{\DPhi{\df-1}(q_I,q_\df)} \; dt_\df }$;
\item
  $ I_M(x_\df) = \sigma \int_{t_I}^{t_F}  { \Braket{q_F|q_\df}        \DPhi{\df-1}(q_\df,q_I)        \; dt_\df }$;
\item
  $ I_R(x_\df) = \sigma \int_{t_F}^{\tMax}{ \Anti{\Braket{q_\df|q_F}} \DPhi{\df-1}(q_\df,q_I)        \; dt_\df }$,
\end{itemize}
and (\ref{eq:psi-bi}) gives us $B_\df$ recursively,
\[
\begin{aligned}
		B_\df
		= \frac{\sigma B_{\df-1}}{\DPhi{\df}(q_F,q_I)}
		&\int_X \left\{ 
					\int_{\tMin}^{t_I}{ \Braket{q_F|q_\df} \Anti{\DPhi{\df-1}(q_I,q_\df)} \; dt_\df } \right. \\ 
		&\qquad \qquad 
				+ \int_{t_I}^{t_F}  { \Braket{q_F|q_\df} \DPhi{\df-1}(q_\df,q_I)        \; dt_\df } \\
		&\qquad \qquad \qquad \qquad
				+ \left. \int_{t_F}^{\tMax}{ \Anti{\Braket{q_\df|q_F}} \DPhi{\df-1}(q_\df,q_I) \; dt_\df }
				\right\} \enspace .
\end{aligned}
\]

\section{Computational Interpretation of the Model}
\label{sec:computational-significance}

To illustrate the full computational significance of our reformulation (especially the bidirectional version), we first need to digress slightly, and explain Eilenberg's (\citeyear{Eil74}) \emph{$X$-machine} model of computation. This is an extremely powerful computational model, which easily captures (and extends) the power of the Turing machine. We will then show that a particle's trajectory can be regarded as an \Xm drawn in spacetime, and that (a minor variant of) this machine computes its own amplitude (as a trajectory).

\subsection{{\Xm}s}
\label{sec:xms}
An \Xm $M = F^\Lambda$ (where $X$ is a data type) is a finite state machine $F$ over some alphabet $A$, together with a \emph{labelling} function $\Lambda \colon a \mapsto a^\Lambda \colon A \to R(X)$, where $R(X)$ is the ring of relations of type $X \leftrightarrow X$.

Each word $w = a_1 \dots a_n$ in the language $\Mod{F}$ recognised by the machine $F$ can be transformed by $\Lambda$ into a relation $w^\Lambda$ on $X$, using the scheme
\[
	w^\Lambda = {a_1}^\Lambda \circ \dots \circ {a_n}^\Lambda
\]
and taking the union of these relations gives the relation $\Mod{F^\Lambda}$ computed by the machine,
\[
	\Mod{F^\Lambda} = \bigcup \Set{ w^\Lambda | w \in \Mod{F} } \enspace .
\]
If we want to model a relation of type $Y \leftrightarrow Z$, for data types $Y \neq Z$, we equip the machine with encoding and decoding relations, $E: Y \to X$ and $D: X \to Z$. Then the behaviour computed by the extended machine is the relation $E \circ \Mod{F^\Lambda} \circ D$.

Although the language $\Mod{F}$ is necessarily regular, the computational power of the \Xm model is unlimited. For, given any set-theoretic relation $\zeta \colon Y \to Z$, we can compute it using the trivial (2-state, 1-transition)-machine with $X = Y \times Z$, by picking any $\Fixed{z} \in Z$, and using the encoder $y^E = \Tuple{y,\Fixed{z}}$, the decoder $\Tuple{y,z}^D = z$, and labelling $a^\Lambda = \overline{\zeta}$, where $\Tuple{y, \Fixed{z}}^{\overline{\zeta}} = \Tuple{y, \zeta(y)}$. For now, given any $y \in Y$, we have $\Mod{F^\Lambda} = \bigcup \Set{a^\Lambda} = \overline{\zeta}$, and
\[
    y^{(E \circ \Mod{F^\Lambda} \circ D)}
  = y^{(E \circ \zeta \circ D)}
  = \Tuple{y, \Fixed{z}}^{(\overline{\zeta} \circ D)}
  = \bigcup \Tuple{y, \zeta(y)}^D
  = \zeta(y) \enspace .
\]

\subsection{Computation by admissible machines}
In our case, all of the path relations we consider will be constant multipliers of the form $k_c \colon z \mapsto zc$, where $c, z \in \Cset$. The resulting machine behaviour will therefore be a set of such multipliers, and we can meaningfully form their sum (which is again a multiplier). For reasons that will shortly become clear, however, we will restrict attention to those paths which visit each state of the machine at least once. We therefore define the \emph{additive behaviour} of such a machine $M = F^\Lambda$ to be the function $\Mod{M}^{+}$ on $\Cset$ given by
\[
	\Mod{M}^{+}(z) = \sum \Set{ w^\Lambda(z) | w \in \Mod{F}, \text{ $w$ visits each state of $F$ at least once }  }
\]
If $M$ is a machine of this form, we will declare the behaviour of $M$ to be the function $\Mod{M}^{+}$, and speak of $M$ as an \emph{additive $X$-machine}. Any finitary path $\Path{q} = q_I \to q_1 \to \dots \to q_\df \to q_F$ generates an additive \Xm $M_{q}$ with state set $\Set{q_I, q_1, \dots, q_\df, q_F}$, alphabet $A = \Set{h_0, \dots, h_\df}$, and transitions $\{ q_n \xrightarrow{h_n} q_{n+1} \;|\; n = 0, \dots, \df \}$. Each transition in the machine is a hop along the path, and is naturally associated with the function ${h_n}^\Lambda = \lambda z . (z . \HA{q_{n+1}|q_n}) : \Cset \to \Cset$ that multiplies any input amplitude $z$ by the hop amplitude $\HA{q_{n+1}|q_n}$. If $M_q$ is an additive \Xm generated by some path $\Path{q}$ with initial state $q_I$, final state $q_F$, and intermediate states in $R$, we shall say that $M$ is \emph{admissible}, and that $\Path{q}$ \emph{generates} $M$. We claim that each path computes its own amplitude, when considered as the machine it generates.

\paragraph{Computation by the unidirectional model.}
For unidirectional machines, each hop $h_n$ involves a jump forward in time, so the states $\Set{q_n}$ must all be distinct, and the path $\Path{q}$ forms a future-pointing chain through spacetime. Consequently, the machine $M_{q}$ recognises precisely one string, and the additive and standard behaviours of the \Xm are identical. The function computed by this path maps each $z \in \Cset$ to
\begin{equation}
	  z^{\left[ ({h_0}^\Lambda) \circ \dots \circ ({h_\df}^\Lambda) \right]}
	= z \times \HA{q_{n+1}|q_n} \HA{q_n|q_{n-1}} \dots \HA{q_1|q_0}
	= z \times \psi\Path{q} \enspace .
	\label{eq:path-comp}
\end{equation}
As claimed, therefore, each (unidirectional) trajectory directly computes its own contribution to the amplitude of any path containing it.

\paragraph{Computation by the bidirectional model.}
Equation (\ref{eq:path-comp}) holds also for unidirectional paths in bidirectional machines, but the general physical interpretation is more complicated, because of the possibility of loops. Essentially, we need to distinguish carefully between two related questions, viz.
\begin{itemize}
\item what is the amplitude that the path $\Path{q}$ is traversed?
\item what is the amplitude that the path $\Path{q}$ is \emph{observed} to have been traversed?
\end{itemize}

To see why, let us suppose that the path $\Path{q}$ contains only one loop, and that $m$ is minimal such that $q_{m+1} = q_{n+1}$ for some $n$ satisfying $m < n$; write the associated sequence of hops as a concatenation of three segments, viz. $h_0 \dots h_\df = u.v.w$, where $u = h_0 \dots h_m$, $v = h_{m+1} \dots h_n$ and $w = h_{n+1} \dots h_\df$. Since $v$ represents a spacetime loop from $q_{m+1}$ back to $q_{n+1} = q_{m+1}$, there is no observable difference between any of the paths $u.v^j.w$, for $j \geq 1$. Consequently, while the amplitude for the path $\Path{q}$ is just $\psi\Path{q}$, the amplitude that this path is \emph{observed} is instead the amplitude $\psi^*\Path{q} = \sum_{j=1}^{\infty}{ \psi\Path{u} \times \left(\psi\Path{v}\right)^j \times \psi\Path{w}}$.

More generally, given the machine $F$ generated by any bidirectional trajectory $\Path{q}$, and any two strings $\alpha$, $\beta$ which are recognised by $F$, \emph{and which visit each state at least once}, there will be no observable difference between $\alpha$ and $\beta$. Consequently, if we define
\[
  F^{+} = \Set{ w^\Lambda | w \in \Mod{F}, \text{ $w$ visits each state at least once }  }
\]
then the amplitude $\psi^{+}$ that $\Path{q}$ is \emph{observed} to have been the path traversed will satisfy, for $z \in \Cset$,
\[
		z . \psi^{+} 
		= \sum \Set{ w^\Lambda(z) | w \in F^{+} }
		= \Mod{F^\Lambda}^{+}(z)
\]
and once again, if we think of $\Path{q}$ as an additive \Xm, it computes its own contribution to the amplitude of any path containing it.

\section{Concluding Arguments}
\label{sec:conclusions}

Recall that an additive \Xm $M$ is \emph{admissible} provided there is some finitary bidirectional path $\Path{q}$ that generates it. Say that two paths $\Path{q}_1$ and $\Path{q}_2$ are \emph{equivalent}, provided they generate precisely the same admissible machine $M$. Clearly, this \emph{is} an equivalence relation, and given any path $\Path{q}$, there will some equivalence class $\widetilde{q}$ containing it. Moreover, the amplitude $\Mod{M}^{+}$ is given by summing the amplitudes of the various paths in $\widetilde{q}$. Consequently, summing over all paths is the same as summing over all admissible machines, so that (regarding $\psi(q_F,q_I)$ as a multiplier),
\[
	\psi(q_F,q_I) = \sum \Set{ \Mod{M}^{+} |  \text{ $M$ is admissible } } \enspace ,
\]
and $\psi(q_F, q_I)$ can be regarded as integrating all of the admissible machine amplitudes. In the bidirectional formulation, then, the nature of motion in quantum theory reveals itself to be inherently computational. It is not that trajectories can be computed; rather, they \emph{are} computations. As a particle hops through spacetime, it simultaneously \emph{constructs} and \emph{executes} a computational state machine, and the amplitude computed by this machine is precisely the amplitude of the trajectory that constructed it.

In section \ref{sec:models:digital-physics}, we noted how digital physics assumes the existence of a computation that computes each universe's history, which suggests that the \Quote{computer} which executes the computation is somehow external to the universes being constructed. In contrast, the bidirectional model is telling us that each universe is a \emph{process}, in which each trajectory is a sub-process which computes its own amplitude. Moreover, all of these sub-processes interact with one another non-locally, because hop amplitudes are based on the classical action, and this in turn depends on the ever-changing spacetime distribution of the other particles. In other words, as we have argued elsewhere, quantum theory is best thought of, not in terms of computation, but in terms of \emph{interactive formal processes} \citep{Sta07}.

Clearly, this idea has echoes of \ItBit, and indeed the bidirectional model helps explain Wheeler's delayed-choice experiment. The apparent paradox relies on two assumptions concerning the experimental set-up. First, the photon must pass through the barrier in order to be observed on the other side; and second, we can reliably identify a time by which the photon has travelled beyond the barrier (we need to make our delayed choice after this time). Both of our reformulations refute the first assumption (the discontinuous nature of hop-based motion means that the Intermediate Value Theorem cannot be invoked to prove that the trajectory necessarily passes through the barrier), while the bidirectional model also refutes the second assumption, since there is no reliable sense in which the decision can be said to have been made \Quote{after} the trajectory intersects the barrier. Thus the delayed-choice experiment contains no paradox, and there is nothing to explain.

We should also be clear as to what our reformulation does \emph{not} say. Throughout this discussion we have focussed on the computational nature of trajectories, but it should be stressed that there is an important distinction to be be drawn between what a process \emph{does}, and how that process is \emph{structured}. This is the same distinction as that highlighted in section \ref{sec:models:digital-physics} between Schmidhuber's and Tegmark's versions of the computational universe hypothesis: whereas Schmidhuber considers process evolutions to be computable, Tegmark requires instead that their descriptions be computable. In our case, while we know that each trajectory computes its amplitude, we cannot say that the amplitude itself is necessarily \Quote{computable} in the Turing sense, because we cannot as yet identify the extent to which the two forms of computation are related. As a \emph{process}, each trajectory is computational, but the \emph{values} it manipulates need not be.

\subsection{Open questions}
\paragraph{(a)} Clearly, we need to determine the relationship between trajectory computations and Turing computations. There must certainly be some such relationship, because the admissible \Xm model underpinning trajectory computation is closely related to the Finite State Machine, which in turn underpins the basic structure of the Turing machine. Are values (like the processes that generate them) constrained to be computable in any standard sense?

\paragraph{(b)} Although we have exchanged continuous motion for motion based on discrete hops, we have not as yet done away with continuous spaces in their entirety, because many of the expressions given in this paper make use of integration. As we argued above, continuity is not directly observable, so we would prefer a purely discrete model. We should therefore investigate the extent to which the formulation presented here can be re-expressed in purely formal terms, for example using the $\pi$-calculus (a standard theoretical vehicle for modelling mobile distributed process-based systems) \citep{Mil99,SW01}. More straightforwardly, can we adapt the models presented here---for example, by replacing integrals with sums---to generate a truly \emph{discrete} models of physics?

\paragraph{(c)} Suppose we impose the condition that whenever a particle hops inside some arbitrary region (which we can think of as the interior of an event horizon), it cannot hop back out again. This will have a global influence upon trajectory amplitudes in the bidirectional model, because every journey would otherwise have had the option to include hops that pass through the excluded region. In particular, the observed positions of geodesics (assuming these can be modelled in terms of finite trajectories?) can be expected to change position, whence the presence of the excluded region will generate a perceived \Quote{warping} of spacetime geometry. Does this warping agree with the warping predicted by, \eg general relativity? Can the bidirectional model be extended to give a model of quantum gravity?

\paragraph{(d)} Feynman's original path-integral methods appear to make various assumptions which we have rejected, including such mainstays of real-world observation as the \emph{arrow of time} and the \emph{continuity of motion}. The status of these assumptions in Feynman's formulation needs, therefore, to be considered in more depth than has been possible here. It may be that they are spurious elements of his construction which play no actual r\^ole, and which are therefore logically independent of his formulae. But if they do indeed play a relevant part in his formulation, they must necessarily become \emph{provable theorems} within both the unidirectional and bidirectional models presented here, because our models agree with Feynman's \emph{by construction}. That is, any property that is (a) expressible in terms of `what is seen by observers', and (b) `built-into' Feynman's assumptions, must necessarily reappear from our own equations, since these give identical results when used to calculate amplitudes.

\theendnotes

\begin{acknowledgements}
This research was supported in part by the EPSRC HyperNet project (Hypercomputation Research Network, grant number EP/E064183/1).
\end{acknowledgements}

\bibliography{UC2008-stannett-springer}

\end{document}